%% file: main.tex
\documentclass[webversion]{usetex-v1}
\usepackage{graphicx}
\usepackage{cite}	
\usepackage{url}
\usepackage{amsfonts}

\long\def\com#1{}

\title{\vspace{-1.5em}Efficient System-Enforced Deterministic Parallelism
	\\ {\tt UNPUBLISHED DRAFT}\vspace{-0.1em}}
\author{
	\authname{Amittai Aviram, Shu-Chun Weng, Sen Hu, Bryan Ford}
	\authaddr{Yale University}
}

\begin{document}

\maketitle

\input{abs}
\input{intro}
\input{kern}

\input{user}

\input{impl}

\input{eval}

\input{related}

\input{concl}

\bibliography{os}
\bibliographystyle{abbr}

\end{document}

%% file: abs.tex
\begin{abstract}
Deterministic execution offers many benefits
for debugging, fault tolerance, and security.
Running {\em parallel} programs deterministically
is usually difficult and costly, however---%
especially if we desire {\em system-enforced} determinism,
ensuring precise repeatability
of arbitrarily buggy or malicious software.
Determinator is a novel operating system
that enforces determinism on both multithreaded and multi-process computations.
Determinator's kernel
provides only single-threaded, ``shared-nothing'' address spaces
interacting via deterministic synchronization.
An untrusted user-level runtime uses distributed computing techniques
to emulate familiar abstractions
such as Unix processes, file systems, and shared memory multithreading.
The system runs parallel applications deterministically
both on multicore PCs and across nodes in a cluster.
Coarse-grained parallel benchmarks
perform and scale comparably to---%
sometimes better than---conventional systems,
though determinism is costly
for fine-grained parallel applications.
\end{abstract}

%% file: intro.tex
\section{Introduction}

It is often useful to run software {\em deterministically},
ensuring a given program and input always yields exactly the same result.
Deterministic execution makes bugs reproducible,
and is required for ``record-and-replay'' debugging~\cite{
	feldman88igor,king05debugging}.
Fault tolerance~\cite{
	schneider90implementing,bressoud96hypervisor,castro99practical}
and accountability mechanisms~\cite{haeberlen07peerreview}
rely on execution being deterministic
and bit-for-bit identical across state replicas.
Intrusion analysis~\cite{dunlap02revirt,joshi05detecting}
and timing channel control~\cite{ford10determinating}
can further benefit from {\em system-enforced determinism},
where the system prevents application code from
depending on execution timing or other unintended inputs
even if the code is maliciously designed to do so.

Multicore processors and ubiquitous parallelism
make programming environments increasingly nondeterministic, however.
Nondeterminism makes software harder to develop and debug~\cite{
	lee06problem,lu08learning}.
Race detectors help~\cite{engler03racerx,musuvathi08heisenbugs},
but even properly synchronized programs may have
higher-level heisenbugs~\cite{artho03high}.
The cost of logging and replaying the internal nondeterministic events
in parallel software~\cite{
	choi98deterministic,dunlap08execution}
can be orders of magnitude higher than that of logging
only a computation's external inputs,
especially for system-enforced replay~\cite{
	dunlap02revirt,dunlap08execution}.
This cost usually precludes logging ``normal-case'' execution,
diminishing the technique's effectiveness.
A heisenbug or intrusion that manifests ``in the field'' with logging disabled
may not reappear during subsequent logged attempts to reproduce it---%
especially with malware {\em designed} to evade analysis
by detecting the timing impact of logging or virtualization~\cite{
	garfinkel07compatibility}.

Motivated by its many uses,
we would like system-enforced determinism to be available
for {\em normal-case} execution of parallel applications.
To test this goal's feasibility,
we built {\em Determinator},
an operating system that not only executes
individual processes deterministically,
as in deterministic user-level scheduling~\cite{
	berger09grace,bergan10coredet},
but can enforce determinism on
hierarchies of interacting processes.
Rerunning a multi-process Determinator computation
with the same inputs yields exactly the same outputs,
without internal event logging.
Determinator treats all potential nondeterministic inputs to a computation---%
including all timing information---%
as ``privileged information,''
which normal applications cannot obtain except via controlled channels.
We treat deterministic execution
as not just a debugging tool but a security principle:
if malware infects an unprivileged Determinator application,
it should be unable to evade replay-based analysis.

System-enforced determinism is challenging
because current programming environments and APIs
are riddled with timing dependencies.
Most shared-memory parallel code uses mutual exclusion primitives:
even when used correctly,
timing determines the application-visible order
in which competing threads acquire a mutex.
Concurrency makes names allocated from shared namespaces,
such as pointers returned by {\tt malloc()}
and file descriptors returned by {\tt open()},
timing-dependent.
Synchronizing operations like semaphores, message queues, and {\tt wait()}
nondeterministically return ``the first'' event, message,
or terminated process available.
Even single-threaded processes are nondeterministic
when run in parallel,
due to their interleaved accesses to shared resources.
A parallel `{\tt make -j}' command often presents
a chaotic mix of its child tasks' outputs,
for example,
and missing dependencies can yield ``makefile heisenbugs''
that manifest only under parallel execution.

Addressing these challenges in Determinator
led us to the insight
that timing dependencies commonly
fall into a few categories:
unintended interactions via shared state or namespaces;
synchronization abstractions with shareable endpoints;
true dependencies on ``real-world'' time;
and application-level scheduling.
Determinator avoids physically shared state
by isolating concurrent activities during normal execution,
allowing interaction only at explicit synchronization points.
The kernel's API uses local, application-chosen names
in place of shared, OS-managed namespaces.
Synchronization primitives operate ``one-to-one,''
between {\em specific} threads,
preventing threads from ``racing'' to an operation.
Determinator treats access to real-world time as I/O, 
controlling it as with other devices such as disk or network.
Finally, Determinator requires scheduling
to be separated from application logic and handled by the system,
or else emulated using a deterministic, virtual notion of ``time.''

Since we wish to derive basic principles
for system-enforced determinism,
Determinator currently makes no attempt
at compatibility with existing operating systems,
and provides limited compatibility with existing APIs.
The kernel's low-level API offers
only one user-visible abstraction, {\em spaces},
representing execution state and virtual memory,
and only three system calls by which spaces synchronize and communicate.
The API's minimality facilitates both experimentation
and reasoning about its determinism.
Despites this simplicity,
our untrusted, user-level runtime builds atop the kernel
to provide familiar programming abstractions.
The runtime uses file replication and versioning~\cite{
	parker83detection}
to offer applications a logically shared file system
via standard APIs;
distributed shared memory~\cite{
	carter91implementation,amza96treadmarks}
to create multithreaded processes logically sharing an address space;
and deterministic scheduling~\cite{
	devietti09dmp,berger09grace,bergan10coredet}
to support pthreads-style synchronization.
Since the kernel enforces determinism,
bugs or vulnerabilities in this runtime
cannot compromise the determinism guarantee.

\com{
While we envision it would be possible to add
multi-process deterministic execution support to existing operating systems,
we decided to take a ``ground-up'' approach
to explore the question of how the goal of system-enforced determinism
might change the properties or constraints 
of an operating system's basic abstractions.
Determinism can also be implemented at application level~\cite{grace,coredet},
or at hardware level in a machine's memory system~\cite{dmp},
but ...
}

Experiments with common parallel benchmarks
suggest that Determinator
can run coarse-grained parallel applications deterministically
with both performance and scalability comparable
to nondeterministic environments.
Determinism incurs a high cost on fine-grained parallel applications, however,
due to Determinator's use of virtual memory to isolate threads.
For ``embarrassingly parallel'' applications
requiring little inter-thread communication,
Determinator can distribute the computation across nodes in a cluster
mostly transparently to the application,
maintaining usable performance and scalability.
The current prototype is merely a proof-of-concept and has many limitations,
such as a restrictive space hierarchy,
limited file system size, no persistent storage,
and inefficient cross-node communication.
Also, our ``clean-slate'' approach is motivated by research goals;
a more realistic approach to deploying system-enforced determinism
would be to add a deterministic ``sandbox''~\cite{
	goldberg96secure,tzicker99integrating}
to a conventional OS.

This paper makes three main contributions.
First, we identify five OS design principles for system-enforced determinism,
and illustrate their application in a novel kernel API.
Second, we demonstrate ways to build familiar abstractions
such as file systems and shared memory
atop a kernel API restricted to deterministic primitives.
Third, we present the first system
that can enforce deterministic execution on multi-process computations
with performance acceptable for ``normal-case'' use,
at least for some (coarse-grained) parallel applications.

Section~\ref{sec-kern} describes
Determinator's kernel design principles and API, then
Section~\ref{sec-user} details
its user-space application runtime.
Section~\ref{sec-impl} examines our prototype implementation, and
Section~\ref{sec-eval} evaluates it informally and experimentally.
Finally,
Section~\ref{sec-related} outlines related work, and
Section~\ref{sec-concl} concludes.

%% file: kern.tex
\section{The Determinator Kernel}
\label{sec-kern}

This section describes Determinator's underlying design principles,
then its low-level execution model and kernel API.
We do not expect normal applications to use the kernel API directly,
but rather the higher-level abstractions the user-level runtime provides,
as described in the next section.
We make no claim that this API is the ``right'' design
for a determinism-enforcing kernel,
but merely use it to explore design challenges and strategies.

\input{princ}

\input{spaces}

\input{api}

\input{reason}
\input{dist}

%% file: princ.tex
\subsection{Kernel API Design Principles}
\label{sec-princ}

We first briefly outline the principles we developed
in designing Determinator,
which address the common sources of timing dependencies we are aware of.
We further discuss the motivations and implications
of these principles below as we detail the kernel API.
We make no claim that this is a complete or conclusive list,
but at least for Determinator these principles prove {\em sufficient}
to offer a deterministic execution guarantee,
for which we briefly sketch formal arguments later in Section~\ref{sec-reason}.

\paragraph{1. Isolate the working state of concurrent activities
	between synchronization points.}
Determinator's kernel API directly provides no shared state abstractions,
such as global file systems or writeable shared memory.
Concurrent activities operate within private ``sandboxes,''
interacting only at deterministically defined synchronization points,
eliminating timing dependencies
due to interleaved access to shared state.

\paragraph{2. Use local, application-chosen names
	instead of global, system-allocated names.}
APIs that assign names from a shared namespace
introduce nondeterminism even when the named objects are unshared:
execution timing affects
the pointers returned by {\tt malloc()} or {\tt mmap()}
or the file numbers returned by {\tt open()}
in multithreaded Unix processes,
and the process IDs returned by {\tt fork()}
or the file names returned by {\tt mktemp()}
in single-threaded processes.
To avoid these sources of nondeterminism,
Determinator's kernel API uses only {\em local} names
chosen by the application:
user-level code decides
where to allocate memory
and what process IDs to assign children.
This principle ensures that
naming a resource reveals no shared state information
other than what the application itself provided.

\paragraph{3. User code determines the participants
	in any synchronization operation,
	and the point in each participant's execution
	at which synchronization occurs.}
The kernel API allows a thread or process
to synchronize with a {\em particular} target,
like Unix processes use {\tt waitpid()}
to wait for a specific child.
The API does {\em not} support
synchronizing with ``any'' or ``the first available'' target
as in Unix's {\tt wait()},
or interrupting another thread at a timing-dependent point in its execution,
as with Unix signals.
Nondeterministic synchronization APIs
may be emulated deterministically, if needed for compatibility,
as described in Section~\ref{sec-dsched}.

\paragraph{4. Treat access to explicit time sources as I/O.}
User code has no direct access to clocks
counting either real time, as in {\tt gettimeofday()},
or nondeterministic ``virtual time'' measures, as in {\tt getrusage()}.
Determinator treats such timing sources as I/O devices
that user code may access only via controlled channels,
as with other devices such as network, disk, and display.

\paragraph{5. Separate application logic from scheduling.}
Deterministic applications cannot make timing-dependent internal
scheduling or load-balancing decisions,
as today's applications often do
using thread pools or work queues.
Applications may {\em expose} arbitrary parallelism
and provide scheduling hints---in principle
they could even download extensions into the kernel
to customize scheduling~\cite{
	bershad95extensibility}---provided
the kernel prevents custom scheduling policies
from affecting computed results.

%% file: spaces.tex
\subsection{Spaces}
\label{sec-spaces}

\begin{figure}[t]
\centering
\includegraphics[width=0.47\textwidth]{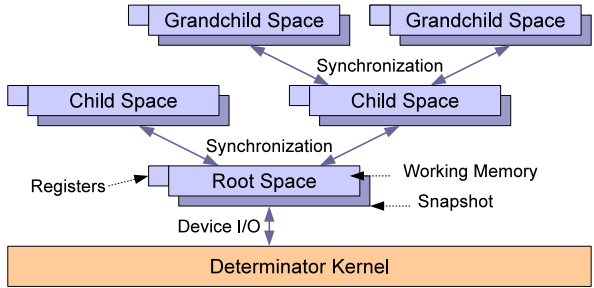}
\caption{The kernel's hierarchy of {\em spaces},
	each containing private register and virtual memory state.}
\label{fig-spaces}
\end{figure}

Determinator executes application code
within a hierarchy of {\em spaces},
illustrated in Figure~\ref{fig-spaces}.
Each space consists of CPU register state for a single control flow,
and private virtual memory containing code and data
directly accessible within that space.
A Determinator space is analogous to a single-threaded Unix process,
with several important differences;
we use the term ``space'' to highlight these differences
and avoid confusion with the
``process'' and ``thread'' abstractions Determinator emulates at user level,
as described later in Section~\ref{sec-user}.

As in a nested process model~\cite{ford96microkernels},
a Determinator space cannot outlive its parent,
and a space can directly interact
{\em only} with its immediate parent and children
via three system calls described below.
Following principle 1 above,
the kernel provides no file systems,
writable shared memory,
or other shared state abstractions.

Following principle 4,
only the distinguished {\em root space}
has direct access to nondeterministic I/O devices including clocks;
other spaces can access I/O devices
only indirectly via parent/child interactions,
or via I/O privileges delegated by the root space.
A parent space can thus control all nondeterministic inputs
into any unprivileged space subtree,
e.g., logging inputs for future replay.
(This space hierarchy also creates a performance bottleneck
for I/O-bound applications,
a limitation of the current design we intend to address in future work.)

\com{
Determinator constrains the inter-process communication and synchronization
of all processes to act as a Kahn process network~\cite{kahn74semantics},
which is provably deterministic despite arbitrary parallelism.
}

\com{	This para is probably confusing and sounds redundant with the above.
The kernel need not enforce strict determinism on all spaces.
The root process has privileges
to access nondeterministic kernel or hardware features, such as timers,
and spaces can selectively delegate these privileges to child spaces.
Once a privileged space creates an unprivileged child space, however,
that child and all of its descendants execute strictly deterministically.
Since this paper focuses on deterministic execution,
we will discuss only the Determinator API features
available to strictly deterministic processes.
}

\com{
\subsection{Execution Model}

A Determinator space can have three states:
{\em runnable}, {\em stopped}, and {\em waiting}.
Runnable processes can execute concurrently with all other runnable processes,
according to a hypervisor-controlled scheduling policy,
but do not interact with each other while running.
A stopped process does nothing until its parent explicitly {\em starts} it.
A waiting process is blocked until a particular child stops,
at which point the waiting process becomes runnable again.
}

%% file: api.tex
\subsection{System Call API}
\label{sec-impl-api}
\label{sec-api}

\begin{table*}[t]
\centering
\begin{small}
\begin{tabular}{c|p{0.90\textwidth}}
Call	& Description \\
\hline
Put 
	& Copy register state and/or a virtual memory range into a child space,
	  and optionally start the child executing. \\
Get	& Copy register state, a virtual memory range,
	  and/or changes since the last snapshot
	  out of a child space. \\
Ret	& Stop and wait for parent to issue a Get or Put. \\
\end{tabular}
\end{small}
\caption{System calls comprising Determinator's kernel API.}
\label{tab-syscalls}
\end{table*}

\begin{table}[t]
\centering
\begin{small}
\begin{tabular}{c|c|c|l}
Put		& Get		& Option & Description \\
\hline
\checkmark	& \checkmark	& Regs	& PUT/GET child's register state. \\
\checkmark	& \checkmark	& Copy	& Copy memory to/from child. \\
\checkmark	& \checkmark	& Zero	& Zero-fill virtual memory range. \\
\checkmark	&		& Snap	& Snapshot child's virtual memory. \\
\checkmark	&		& Start	& Start child space executing. \\
		& \checkmark	& Merge	& Merge child's changes into parent. \\
\checkmark	& \checkmark	& Perm	& Set memory access permissions. \\
\checkmark	& \checkmark	& Tree	& Copy (grand)child subtree. \\
\end{tabular}
\end{small}
\caption{Options/arguments to the Put and Get calls.}
\label{tab-sysopts}
\end{table}

Determinator spaces interact only as a result of processor traps
and the kernel's three system calls---Put, Get, and Ret,
summarized in Table~\ref{tab-syscalls}.
Put and Get take several optional arguments,
summarized in Table~\ref{tab-sysopts}.
Most options can be combined:
e.g., in one Put call a space can initialize a child's registers,
copy a range of the parent's virtual memory into the child,
set page permissions on the destination range,
save a complete snapshot of the child's address space,
and start the child executing.

As per principle 2 above,
each space has a private namespace of child spaces,
which user-level code manages.
A space specifies a child number to Get or Put,
and the kernel creates that child if it doesn't already exist,
before performing the requested operations.
If the specified child did exist and was still executing
at the time of the Put/Get call,
the kernel blocks the parent's execution until the child stops
due to a Ret system call or a processor trap.
These ``rendezvous'' semantics
ensure that spaces synchronize only at well-defined points
in both spaces' execution,
as required by principle 3.

The Copy option logically copies
a range of virtual memory
between the invoking space and the specified child.
The kernel uses copy-on-write
to optimize large copies and avoid physically copying read-only pages.

Merge is available only on Get calls.
A Merge is like a Copy,
except the kernel copies only bytes that {\em differ}
between the child's current and reference snapshots
into the parent space,
leaving other bytes in the parent untouched.
The kernel also detects conflicts:
if a byte changed in {\em both}
the child's and parent's spaces since the snapshot,
the kernel generates an exception,
treating a conflict as a programming error
like an illegal memory access or divide-by-zero.
Determinator's user-level runtime uses Merge
to give multithreaded processes the illusion of shared memory,
as described later in Section~\ref{sec-threads}.
In principle, user-level code could implement Merge itself,
but the kernel's direct access to page tables
makes it easy for the kernel to implement Merge efficiently.

Finally,
the Ret system call stops the calling space,
returning control to the space's parent.
Exceptions such as divide-by-zero also cause a Ret,
providing the parent a status code
indicating why the child stopped.

To facilitate debugging
and prevent untrusted children from looping forever,
a parent can start a child with an {\em instruction limit},
forcing control back to the parent
after the child and its descendants
collectively execute this many instructions.
Counting instructions instead of ``real time''
preserves determinism,
while enabling spaces to ``quantize'' a child's execution
to implement scheduling schemes deterministically at user level~\cite{
	devietti09dmp,bergan10coredet}.
\com{
To implement this feature,
the hypervisor must 
Some processor architectures natively support
control recovery after a precise instruction count~\cite{hp94parisc},
but this capability can be simulated on the x86~\cite{dunlap02revirt}.
}

%% file: reason.tex
\subsection{Reasoning about Determinism}
\label{sec-reason}

Can we be certain the kernel API above indeed guarantees
that space subtrees execute deterministically despite parallelism?
While a detailed proof is out of scope,
we briefly sketch two formal arguments for this guarantee.

The first argument leverages an existing formal parallel computing model:
a Kahn process network~\cite{kahn74semantics}
is a network of single-threaded processes,
which run sequential code deterministically
and interact only via blocking, one-to-one message channels.
Under these restrictions,
a Kahn network behaves deterministically.
Determinator's Get, Put, and Ret calls
are implementable in terms of messages on one-to-one channels,
making Determinator's space hierarchy
formally equivalent to a Kahn process network,
thereby ensuring its determinism.

For a more ``first-principles'' argument, consider a graph
of possible execution traces of a space hierarchy.
Each node represents a synchronization point
in a possible execution history of one space,
vertical edges represent local computation sequences in one space
between synchronization points,
and horizontal edges represent pairwise interactions
where a parent space's Get or Put synchronizes with a child's Ret.
From this graph we construct a ``happens-before'' partial order
over all synchronization points in all possible executions.
At each synchronization point,
assuming all prior (on the partial order)
computation sequences and synchronization interactions
yield only one possible result for a given set of inputs,
then the same is true after that synchronization point:
each synchronization point in a parent space
interacts with only one corresponding point
in a specific child, and vice versa,
and synchronization effects such as memory changes
depend only on the two spaces' states prior to synchronization.
By induction on the partial order,
the entire execution history is therefore deterministic.

%% file: dist.tex
\subsection{Distribution via Space Migration}
\label{sec-dist}

The kernel allows space hierarchies to span
not only multiple CPUs in a multiprocessor/multicore system,
but also multiple nodes in a cluster,
mostly transparently to application code.
While distribution is semantically transparent to applications,
we say ``mostly transparently'' because
an application may have to be designed with distribution in mind
to achieve acceptable performance.
As with other aspects of the kernel's design,
we make no pretense that this is the ``right'' approach
to cross-node distribution,
but merely one way to extend a deterministic execution model
across a cluster.

\begin{figure}[t]
\centering
\includegraphics[width=0.47\textwidth]{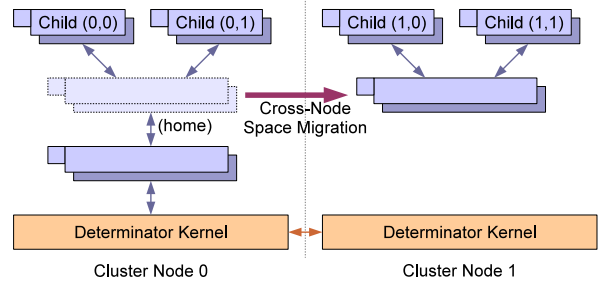}
\caption{A spaces migrating among two nodes
	and starting child spaces on each node.}
\label{fig-migr}
\end{figure}

Distribution support
adds no new system calls or options to the API above.
Instead,
the Determinator kernel interprets the higher-order bits
in each process's child number namespace as a ``node number'' field.
When a space invokes Put or Get,
the kernel first logically migrates
the calling space's state and control flow
to the node whose number the user specifies as part of its child number argument,
before creating and/or interacting with a child on that node
specified in the remaining child number bits.
Figure~\ref{fig-migr} illustrates a space migrating between two nodes
and managing child spaces on each.

Once created,
a space has a {\em home node},
to which the space migrates 
when interacting with its parent on a Ret or trap.
Nodes are numbered so that ``node zero''
in any space's child namespace
always refers to the space's home node.
If a space uses only the low bits in its child numbers
and leaves the node number field zero,
the space's children all have the same home as the parent.

When the kernel migrates a space,
it first transfers to the receiving kernel
only the space's register state and address space summary information.
Next, the receiving kernel requests the space's memory pages
on demand as the space accesses them on the new node.
Each node's kernel avoids redundant cross-node page copying
in the common case when a space repeatedly migrates among several 
nodes---e.g., when a space starts children on each of several nodes,
then returns later to collect their results.
For pages that the migrating space only reads and never writes,
such as program code,
each kernel reuses cached copies of these pages
whenever the space returns to that node.
The kernel currently performs no prefetching or other adaptive optimizations.
Its rudimentary messaging protocol
runs directly atop Ethernet,
and does not support TCP/IP for Internet-wide distribution.

%% file: user.tex
\section{Emulating High-Level Abstractions}
\label{sec-user}

The kernel API described above
eliminates many conveniences to which developers and users are accustomed.
Can we reproduce them under the constraint of strict determinism?
We find that many familiar abstractions remain feasible,
although some semantically nondeterministic abstractions
may be costly to emulate precisely.
This section details the user-level runtime infrastructure
we developed to emulate traditional Unix processes, file systems,
threads, and synchronization under Determinator.

\input{procs}

\input{fs}

\input{io}

\input{threads}

\input{dsched}

%% file: procs.tex
\subsection{Processes and fork/exec/wait}
\label{sec-procs}

We make no attempt to replicate Unix process semantics exactly,
but would like to emulate
traditional {\tt fork}/{\tt exec}/{\tt wait} APIs enough
to support common uses in scriptable shells, build tools,
and multi-process ``batch processing'' applications such as compilers.

\paragraph{\bf Fork:}
Implementing a basic Unix {\tt fork()}
requires only one Put system call,
to copy the parent's entire memory state into a child space,
set up the child's registers,
and start the child.
The difficulty arises from Unix's global process ID (PID) namespace,
a source of nondeterminism
violating our design principle 2 (Section~\ref{sec-princ}).
Since most applications use
PIDs returned by {\tt fork()}
merely as an opaque argument to a subsequent {\tt waitpid()},
our runtime makes PIDs local to each process:
one process's PIDs are unrelated to,
and may numerically conflict with,
PIDs in other processes.
This change breaks Unix applications
that pass PIDs among processes,
and means that commands like `{\tt ps}'
must be built into shells
for the same reason that `{\tt cd}' already is.
This simple approach
works for compute-oriented applications
following the typical fork/wait pattern,
however.

Since {\tt fork()} returns a PID chosen by the system,
while our kernel API requires user code to manage child numbers,
our user-level runtime
maintains a ``free list'' of child spaces
and reserves one during each {\tt fork()}.
To emulate Unix process semantics more closely,
a central space such as the root space
could manage a global PID namespace,
at the cost of requiring inter-space communication
during operations such as {\tt fork()}.

\paragraph{\bf Exec:}
A user-level implementation of Unix {\tt exec()}
must construct the new program's memory image,
intended to replace the old program,
while still executing the old program's runtime library code.
Our runtime loads the new program
into a ``reserved'' child space never used by {\tt fork()},
then calls Get to copy that child's entire memory
atop that of the (running) parent:
this Get thus ``returns'' into the new program.
To ensure that the instruction address
following the old program's Get
is a valid place to start the new program,
the runtime places this Get in a small ``trampoline'' code fragment
mapped at the same location in the old and new programs.
The runtime also carries over some Unix process state,
such as the the PID namespace and file system state described later,
from the old to the new program.

\paragraph{\bf Wait:}
When an application calls {\tt waitpid()}
to wait for a specific child,
the runtime calls Get to synchronize with the child's Ret
and obtain the child's exit status.
(The child may return to the parent
before it wishes to terminate,
in order to make I/O requests as described below;
in this case, the parent's runtime
services the I/O request
and resumes the {\tt waitpid()}
transparently to the application.)

Unix's {\tt wait()} is more challenging,
as it violates principle 3
by waiting for {\em any} (i.e., ``the first'') child to terminate.
Our kernel's API provides no system call to ``wait for any child,''
and can't (for unprivileged spaces)
without violating its determinism guarantee.
Instead, our runtime waits for
the child that was forked earliest
whose status was not yet collected.
This behavior does not affect applications
that fork one or more children
and then wait for all of them to complete,
but affects two common uses of {\tt wait()}.
First, interactive Unix shells use {\tt wait()}
to report when background processes complete;
thus, an interactive shell running under Determinator
requires special ``nondeterminism privileges''
to provide this functionality
(and related functions such as interactive job control).
Second, our runtime's behavior may adversely affect
the performance of programs that use {\tt wait()}
to implement dynamic scheduling or load balancing in user space,
which violates principle 5.

\begin{figure}[t]
\centering
\includegraphics[width=0.47\textwidth]{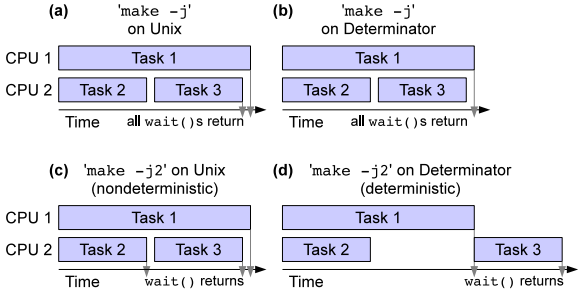}
\caption{Example parallel {\tt make} scheduling scenarios
	under Unix versus Determinator:
	(a) and (b) with unlimited parallelism (no user-level scheduling);
	(c) and (d) with a ``2-worker'' quota imposed at user level.}
\label{fig-pmake}
\end{figure}

Consider a parallel {\tt make}
run with or without limiting
the number of concurrent children.
A plain `{\tt make -j}', allowing unlimited children,
leaves scheduling decisions to the system.
Under Unix or Determinator,
the kernel's scheduler dynamically assigns tasks to available CPUs,
as illustrated in Figure~\ref{fig-pmake} (a) and (b).
If the user runs `{\tt make -j2}', however,
then {\tt make} initially starts only tasks 1 and 2,
then waits for one of them to complete before starting task 3.
Under Unix,
{\tt wait()} returns when the short task 2 completes,
enabling {\tt make} to start task 3 immediately as in (c).
On Determinator, however, the {\tt wait()} returns
only when (deterministically chosen) task 1 completes,
resulting in a non-optimal schedule (d):
determinism prevents the runtime
from learning which of tasks 1 and 2 completed first.
This example illustrates the importance
of separating scheduling from application logic,
as per principle 5.

%% file: fs.tex
\subsection{A Shared File System}
\label{sec-fs}

Unix's globally shared file system provides
a convenient namespace and repository
for staging program inputs, storing outputs,
and holding intermediate results such as temporary files.
Since our kernel permits no physical state sharing,
user-level code must emulate shared state abstractions.
Determinator's ``shared-nothing'' space hierarchy
is similar to a distributed system consisting only of uniprocessor machines,
so our user-level runtime borrows distributed file system principles
to offer applications a shared file system abstraction.

Since our current focus is on emulating familiar abstractions
and not on developing storage systems,
Determinator's file system currently provides no persistence:
it effectively serves only as a temporary file system.
\com{
Later in Section~\ref{sec-future} we will briefly outline
potential extensions for persistent storage.
}

\begin{figure}[t]
\centering
\includegraphics[width=0.30\textwidth]{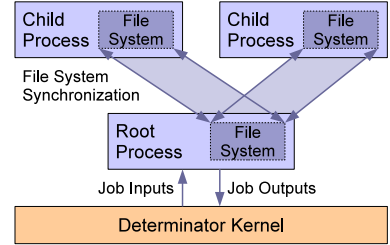}
\caption{Each process's user-level runtime maintains an individual replica
	of a logically shared file system,
	using file versioning to reconcile replicas at synchronization points.}
\label{fig-fs}
\end{figure}

While many distributed file system designs may be applicable,
our runtime uses replication with weak consistency~\cite{
	walker83locus,terry95managing}.
Our runtime
maintains a complete file system replica
in the address space of each process it manages,
as shown in Figure~\ref{fig-fs}.
When a process creates a child via {\tt fork()},
the child inherits a copy of the parent's file system
in addition to the parent's open file descriptors.
Individual {\tt open}/\linebreak[0]{\tt close}/%
\linebreak[0]{\tt read}/\linebreak[0]{\tt write} operations
in a process use only that process's file system replica,
so different processes' replicas may diverge
as they modify files concurrently.
When a child terminates and its parent collects its state via {\tt wait()},
the parent's runtime
copies the child's file system image into a scratch area in the parent space
and uses file versioning~\cite{parker83detection}
to propagate the child's changes into the parent.

If a shell or parallel {\tt make} forks several compiler processes in parallel,
for example,
each child writes its output {\tt .o} file to its own file system replica,
then the parent's runtime merges the resulting {\tt .o} files
into the parent's file system
as the parent collects each child's exit status.
This copying and reconciliation
is not as inefficient as it may appear,
due to the kernel's copy-on-write optimizations.
Replicating a file system image among many spaces
copies no physical pages until user-level code modifies them,
so all processes' copies of identical files
consume only one set of pages.

As in any weakly-consistent file system,
processes may cause {\em conflicts}
if they perform unsynchronized, concurrent writes to the same file.
When our runtime detects a conflict,
it simply discards one copy
and sets a conflict flag on the file;
subsequent attempts to {\tt open()} the file result in errors.
This behavior is intended for batch compute applications
for which conflicts indicate an application or build system bug,
whose appropriate solution is to fix the bug and re-run the job.
Interactive use would demand a conflict handling policy
that avoids losing data.
The user-level runtime could alternatively
use pessimistic locking
to implement stronger consistency
and avoid unsynchronized concurrent writes,
at the cost of more inter-space communication.

The current design's placement of each process's file system replica
in the process's own address space has two drawbacks.
First, it limits total file system size to less than
the size of an address space;
this is a serious limitation in our 32-bit prototype,
though it may be less of an issue on a 64-bit architecture.
Second, wild pointer writes in a buggy process may corrupt the file system
more easily than in Unix,
where a buggy process must actually call {\tt write()} to corrupt a file.
The runtime could address the second issue
by write-protecting the file system area between calls to {\tt write()},
or it could address both issues by storing file system data
in child spaces not used for executing child processes.

%% file: io.tex
\subsection{Input/Output and Logging}
\label{sec-io}

Since unprivileged spaces can access external I/O devices
only indirectly via parent/child interaction within the space hierarchy,
our user-level runtime treats I/O
as a special case of file system synchronization.
In addition to regular files,
a process's file system image can contain special {\em I/O files},
such as a console input file and a console output file.
Unlike Unix device special files,
Determinator's I/O files actually hold data in the process's file system image:
for example, a process's console input file accumulates
all the characters the process has received from the console,
and its console output file contains
all the characters it has written to the console.

When a process does a {\tt read()} from the console,
the C library first returns unread data
already in the process's local console input file.
When no more data is available,
instead of returning an end-of-file condition,
the process calls Ret to synchronize with its parent
and wait for more console input
(or in principle any other form of new input)
to become available.
When the parent does a {\tt wait()}
or otherwise synchronizes with the child,
it propagates any new input it already has to the child.
When the parent has no new input for any waiting children,
it forwards all their input requests to its parent,
and ultimately to the kernel via the root process.

When a process does a console {\tt write()},
the runtime appends the new data to its internal console output file
as it would append to a regular file.
The next time the process synchronizes with its parent,
file system reconciliation
propagates these writes toward the root process,
which forwards them to the kernel's I/O devices.
A process can request immediate synchronization and output propagation
by explicitly calling {\tt fsync()}.

The file system reconciliation mechanism handles ``append-only'' writes
differently from other file changes,
enabling processes to write concurrently to the console or to log files
without conflict.
During reconciliation, if both the parent and child process
have made append-only writes to the same file,
reconciliation appends the child's latest writes
to the parent's copy of the file,
and appends the parent's latest writes to the child's copy.
Each process's output file thus accumulates all processes' concurrent writes,
though different processes may observe these writes in a different order.
\com{
In the common case where the synchronization event is the child's termination,
this ordering variation never becomes visible
because the terminated child's copy of the output file is discarded.
This ordering variation could be visible to processes
that synchronize with their parents before termination;
the runtime could enforce a total ordering if needed
by coordinating with some central ``ordering manager'' such as the root space,
at the cost of additional synchronization.
}
Unlike Unix,
rerunning a parallel computation from the same inputs
with and without output redirection
yields byte-for-byte identical console and log file output.

%% file: threads.tex
\subsection{Shared Memory Multithreading}
\label{sec-threads}

Shared memory multithreading is popular
despite the nondeterminism it introduces into processes,
in part because parallel code need not pack and unpack messages:
threads simply compute ``in-place''
on shared variables and structures.
Since Determinator gives user spaces
no physically shared memory
other than read-only sharing via copy-on-write,
emulating shared memory
involves distributed shared memory (DSM) techniques.

As with file systems, there are many approaches to DSM,
but ours builds on release-consistent DSM~\cite{
	carter91implementation,amza96treadmarks},
which balances efficiency with programming convenience.
Although release consistency normally makes memory access behavior
even {\em less} deterministic by relaxing the rules of sequential consistency,
we have adapted it into a memory model
we call {\em deterministic consistency} (DC),
which we detail elsewhere~\cite{
	ford10deterministic}.
DC's roots lie in early parallel Fortran systems~\cite{
	schwartz80burroughs,beltrametti88control},
in which all processors make private copies of shared data
at the beginning of a parallel ``for'' loop,
then read and modify only their private ``workspaces'' within the loop,
and merge their results once all processors complete.

DC propagates memory changes between threads predictably,
only at program-defined synchronization points.
If one thread executes the assignment `$x = y$'
while another concurrently executes `$y = x$',
for example,
this code yields a nondeterministic data race in standard memory models,
but in DC it is race-free and always swaps $x$ with $y$.
DC's semantics might simplify simulations
in which threads running in lock-step
read and update large arrays in-place, for example.
The absence of read/write conflicts in DC also simplifies implementation,
eliminating the need to execute parallel sequences speculatively
and risk aborting and wasting effort if a dependency is detected,
as when deterministically emulating sequential consistency~\cite{
	devietti09dmp,berger09grace,bergan10coredet}.

\begin{figure}[t]
\centering
\includegraphics[width=0.47\textwidth]{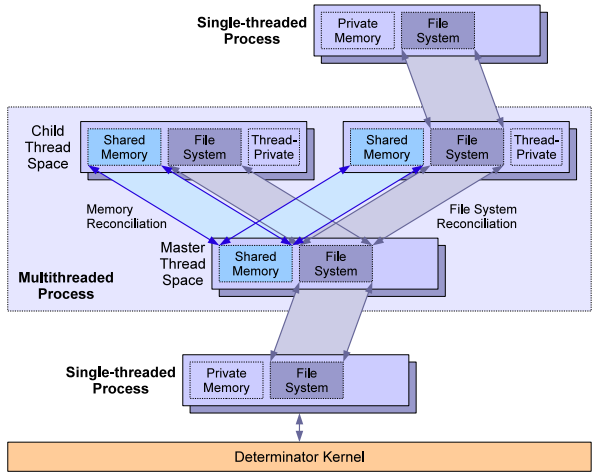}
\caption{A multithreaded process built from one space per thread,
	with a master space
	managing synchronization and memory reconciliation.}
\label{fig-threads}
\end{figure}

Our runtime uses the kernel's Snap and Merge operations
(Section~\ref{sec-api})
to emulate shared memory with deterministic consistency
and ``fork/join'' thread synchronization.
To fork a child, the parent thread
calls Put with the Copy, Snap, Regs, and Start options
to copy the shared part of its memory into a child space,
save a snapshot of that memory state in the child,
and start the child running,
as illustrated in Figure~\ref{fig-threads}.
The master thread may fork multiple children in parallel this way.
To synchronize with a child and collect its results,
the parent calls Get with the Merge option,
which merges all changes the child made to its shared address space,
since the child's snapshot was taken, back into the parent's space.
If both the parent and child---%
or the child and other children whose changes the parent has collected---%
have concurrently modified the same shared memory byte since the snapshot,
the kernel detects and reports this write/write conflict
(which is DC's only form of data race).

Our runtime also supports barriers,
the foundation of data-parallel programming models like OpenMP~\cite{openmp08}.
When each thread in a group arrives at a barrier,
it calls Ret to stop and wait for the parent thread managing the group.
The parent calls Get with Merge
to collect each child's changes before the barrier,
then calls Put with Copy and Snap to resume each child
with a new shared memory snapshot containing all threads' prior results.
While DC conceptually extends to
non-hierarchical synchronization patterns as well~\cite{ford10deterministic},
such as Lisp-style futures~\cite{halstead85multilisp},
our kernel's current strict space hierarchy
naturally supports only hierarchical synchronization,
a limitation we intend to address in the future.
{\em Any} synchronization abstraction may be emulated at some cost
as described in the next section, however.

An application can choose which parts of its address space
to share and which to keep thread-private.
By placing thread stacks outside the shared region,
all threads can reuse the same stack area,
and the kernel wastes no effort merging stack data.
If threads wish to pass pointers to stack-allocated structures, however,
then they may locate their stacks in disjoint shared regions.
Similarly, if the file system area is shared,
then the threads share a common file descriptor namespace
as in Unix.
Excluding the file system area from shared space
and using normal file system reconciliation
(Section~\ref{sec-fs}) to synchronize it
yields thread-private file tables.

\begin{figure}[t]
\begin{small}
\begin{tabbing}
{\bf md5search}(unsigned char *{\em hash}, int {\em len}, int {\em nthreads}) \\
~~~ \=	char {\em buf}[{\em len}+1], {\em output}[{\em len}+1]; \\
\>	int {\em done} = 0, {\em found} = 0, {\em i}; \\
\>	{\bf first\_string}(\&{\em buf}, {\em len}); \\
\>	while \= (!{\em done} \&\& !{\em found}) \\
\>	\>	for \= ({\em i} = 0; {\em i} $<$ {\em nthreads}; {\em i}++) \\
\>	\>	\>	{\bf next\_string}(\&{\em buf}, {\em len}, \&{\em done}); \\
\>	\>	\>	if \= ({\bf thread\_fork}({\em i}) == IN\_CHILD) \\
\>	\>	\>	\>	{\bf check\_md5}(\&{\em buf}, {\em hash}, \&{\em output}, \&{\em found}); \\
\>	\>	\>	\>	{\bf thread\_exit}(); \\
\>	\>	for \= ({\em i} = 0; {\em i} $<$ {\em nthreads}; {\em i}++) \\
\>	\>	\>	{\bf thread\_join}({\em i});
\end{tabbing}
\end{small}
\caption{Pseudocode for parallel ``MD5 cracker.''}
\label{fig-md5code}
\end{figure}

The C pseudocode in Figure~\ref{fig-md5code},
a simplified fragment of a brute-force ``MD5 cracking'' benchmark
we use later in Section~\ref{sec-eval},
illustrates two convenient properties of deterministic consistency.
First,
since threads can have private stacks in overlapping address ranges,
{\tt thread\_fork()} acts like Unix's process-level {\tt fork()},
cloning the parent's stack into the child,
so the program need not separate the child thread's code
into a separate function as pthreads requires.
Second,
the parent thread's {\tt next\_string()} call updates {\em buf} in-place
before forking each child,
whose ``work function'' {\tt check\_md5()} refers to this buffer.
In a nondeterministic thread model,
this code contains a data race:
the parent may update {\em buf} for the next child
before the previous child has finished reading it.
Under Determinator, however, this code is race-free:
each child's view of {\em buf} remains as it was when that child was forked,
until the child explicitly calls {\tt thread\_exit()}.

%% file: dsched.tex
\subsection{Legacy Synchronization APIs}
\label{sec-dsched}

Although some synchronization abstractions
naturally fit a deterministic model,
others do not.
Mutex locks are semantically nondeterministic:
that they guarantee that only one thread may own a lock at once,
but allow competing threads to acquire the lock in any order.
Condition variables, semaphores, and message queues
allow multiple threads to race to signal, post, or send, respectively,
and these events wake up any of several waiting or reading threads,
violating our principle 3.

For existing sequential code not yet parallelized,
we hope this code might be parallelized
using naturally deterministic synchronization abstractions
like data-parallel programming models such as OpenMP~\cite{openmp08}
and SHIM~\cite{edwards08programming} provide.
For code already parallelized using nondeterministic synchronization, however,
Determinator's runtime can emulate the standard pthreads API
via deterministic scheduling~\cite{
	devietti09dmp,berger09grace,bergan10coredet},
at certain costs.

In a process that uses nondeterministic synchronization,
the process's initial {\em master space}
never runs application code directly,
but instead runs a {\em deterministic scheduler}.
This scheduler creates one child space to run each application thread.
The scheduler runs the threads under an artificial execution schedule,
emulating a schedule by which a true shared-memory multiprocessor
might in principle run them,
but using a deterministic, virtual notion of ``time''---%
e.g., number of instructions executed---%
to schedule thread interactions.

Like DMP~\cite{devietti09dmp,bergan10coredet},
our deterministic scheduler {\em quantizes} each thread's execution
by preempting it after executing a fixed number of instructions.
Whereas DMP implements preemption by instrumenting user-level code,
our scheduler uses
the kernel's instruction limit feature (Section~\ref{sec-api}).
The scheduler ``donates'' execution quanta to threads round-robin,
allowing each thread to run concurrently with other threads for one quantum,
before collecting the thread's shared memory changes via Merge
and restarting it for another quantum.

A thread's shared memory writes propagate to other threads
only at the end of each quantum,
violating sequential consistency~\cite{lamport79multi}.
Like DMP-B~\cite{bergan10coredet},
our deterministic scheduler implements
release consistency~\cite{gharachorloo90memory},
totally ordering only synchronization operations.
To enforce this total order,
each synchronization operation could simply spin for a a full quantum.
To avoid wasteful spinning, however,
our synchronization primitives
interact with the deterministic scheduler directly.

Each mutex, for example, is always ``owned'' by some thread,
whether or not the mutex is locked.
The mutex's owner can lock and unlock the mutex
without scheduler interactions,
but any other thread needing the mutex
must first invoke the scheduler to obtain ownership.
At the current owner's next quantum,
the scheduler ``steals'' the mutex from its current owner
if the mutex is unlocked,
and otherwise places the locking thread on the mutex's queue
to be awoken once the mutex is available.

Since the scheduler can preempt threads at any point,
a challenge common to any preemptive scenario
is making synchronization functions such as \verb|pthread_mutex_lock()| atomic.
The kernel does not allow threads
to disable or extend their own instruction limits,
since we wish to use instruction limits at process level as well,
e.g., to enforce deterministic ``time'' quotas on untrusted processes,
or to improve user-level process scheduling (see Section~\ref{sec-procs})
by quantizing process execution.
After synchronizing with a child thread, therefore,
the master space checks whether the instruction limit
preempted a synchronization function,
and if so,
resumes the preempted code in the master space.
Before returning to the application,
these functions check whether they have been ``promoted'' to the master space,
and if so migrate their register state back to the child thread
and restart the scheduler in the master space.

While deterministic scheduling
provides compatibility with existing parallel code,
it has drawbacks.
The master space,
required to enforce a total order on synchronization operations,
may be a scaling bottleneck
unless execution quanta are large.
Since threads can interact only at quanta boundaries, however,
large quanta increase the time one thread may waste waiting for another,
to steal an unlocked mutex for example.

Further, since the deterministic scheduler may preempt a thread
and propagate shared memory changes at any point in application code,
the {\em programming model} remains nondeterministic.
If one thread runs `$x=y$' while another runs `$y=x$',
the result may be {\em repeatable}
but is no more {\em predictable} to the programmer
than on traditional systems---%
in contrast with the previous section's multithreading model.
While rerunning a program with {\em exactly} identical inputs
will yield identical results,
if the input is perturbed
to change the length of any instruction sequence,
these changes may cascade into a different execution schedule
and trigger {\em schedule-dependent}
if not timing-dependent heisenbugs.

\com{
\subsection{Resource Management}

What if a job runs too long or forever?
Customer can set a time limit or instruction limit or kill it remotely.
If the limit is a deterministic quantum,
then customer can get back the partial execution state
after expiration,
for debugging etc.
}

%% file: impl.tex
\section{Prototype Implementation}
\label{sec-impl}

Determinator is implemented in C with small assembly fragments,
runs on the 32-bit x86 architecture,
and implements the kernel API
and user-level runtime facilities described above.
Source code is available on request.

Since our focus is on parallel compute-bound applications,
Determinator's I/O capabilities are currently limited.
The system provides text-based console I/O
and a Unix-style shell supporting redirection
and both scripted and interactive use.
The shell offers no interactive job control,
which would require currently unimplemented
``nondeterministic privileges'' (Section~\ref{sec-procs}).
The system has no demand paging or persistent disk storage:
the user-level runtime's logically shared file system abstraction
currently operates in physical memory only.

The kernel supports application-transparent space migration
among up to 32 machines in a cluster,
as described in Section~\ref{sec-dist}.
Migration uses a synchronous messaging protocol
with only two request/response types
and implements almost no optimizations such as page prefetching.
The protocol runs directly atop Ethernet,
and is not intended for Internet-wide distribution.

Implementing instruction limits (Section~\ref{sec-api})
requires the kernel to recover control
after a precise number of instructions execute in user mode.
While the PA-RISC architecture provided this feature~\cite{hp94parisc},
the x86 does not,
so we borrowed ReVirt's technique~\cite{dunlap02revirt}.
We first set an {\em imprecise} hardware performance counter,
which unpredictably overshoots its target a small amount,
to interrupt the CPU before the desired number of instructions,
then run the remaining instructions under debug tracing.
\com{
so we used a technique inspired by ReVirt~\cite{dunlap02revirt}.
The kernel first sets a hardware performance counter~\cite{intel09}
to cause an {\em imprecise} interrupt
after retiring fewer than the number of instructions required,
compensating for the unpredictable number of instructions
the counter may ``overshoot'' before the interrupt is delivered;
the kernel then uses the processor's trace flag (TF)
to single-step through the remaining count.
This single-stepping is inefficient,
and could be improved by scanning user code
and stepping one branch at a time, as ReVirt does.
}

%% file: eval.tex
\section{Evaluation}
\label{sec-eval}

This section evaluates the Determinator prototype,
first informally,
then examining single-node and distributed parallel processing performance,
and finally code size.

\subsection{Experience Using the System}

We find that a deterministic programming model
simplifies debugging of both applications and user-level runtime code,
since user-space bugs are always reproducible.
Conversely, when we do observe nondeterministic behavior,
it can result only from a kernel (or hardware) bug,
immediately limiting the search space.

Because Determinator's file system
holds a process's output until the next synchronization event
(often the process's termination),
each process's output appears as a unit
even if the process executes in parallel
with other output-generating processes.
Further, different processes' outputs
appear in a consistent order across runs,
as if run sequentially.
(The kernel provides a system call for debugging
that outputs a line to the ``real'' console immediately,
reflecting true execution order,
but chaotically interleaving output like standard systems.)

While race detection tools exist~\cite{
	engler03racerx,musuvathi08heisenbugs},
we found it convenient that Determinator
detects races all the time under ``normal-case'' execution,
without requiring the user to run a special tool.
Since the kernel detects shared memory conflicts
and the user-level runtime detects file system conflicts
at every synchronization event,
Determinator's model makes race detection
as standard as detecting division by zero or illegal memory accesses.

A subset of Determinator doubles as {\em PIOS},
``Parallel Instructional Operating System,''
which we used in Yale's operating system course this spring.
While the OS course's objectives did not include determinism,
they included introducing students
to parallel, multicore, and distributed operating system concepts.
For this purpose, we found Determinator/PIOS to be a useful instructional tool
due to its simple design, minimal kernel API,
and adoption of distributed systems techniques
within and across physical machines.
PIOS is partly derived from MIT's JOS~\cite{mit-jos},
and includes a similar instructional framework
where students fill in missing pieces of a ``skeleton.''
The twelve students who took the course,
working in groups of two or three,
all successfully reimplemented Determinator's core features:
multiprocessor scheduling with Get/Put/Ret coordination,
virtual memory with copy-on-write and Snap/Merge,
user-level threads with fork/join synchronization
(but not deterministic scheduling),
the user-space file system with versioning and reconciliation,
and application-transparent cross-node distribution via space migration.
In their final projects they extended the OS
with features such as graphics, pipes, and a remote shell.
While instructional use
is by no means indicates a system's real-world utility,
we find the success of the students
in understanding and building on Determinator's architecture promising.

\com{
One of the authors of this paper was a student taking the course who,
initially unware of our intention to write this paper,
chose to refine and evaluate his implementation of cross-node migration
as his final course project,
leading to some of the results presented below.
}

\subsection{Single-node Multicore Performance}

Since Determinator runs user-level code ``natively'' on the hardware
instead of rewriting user code~\cite{
	devietti09dmp,bergan10coredet},
we expect it to perform comparably to conventional systems
when executing single-threaded, compute-bound code.
Since space interactions require system calls,
context switches, and virtual memory operations, however,
we expect determinism to incur a performance cost
in proportion to the amount of interaction between spaces.

\begin{figure}[t]
\centering
\includegraphics[width=0.48\textwidth]{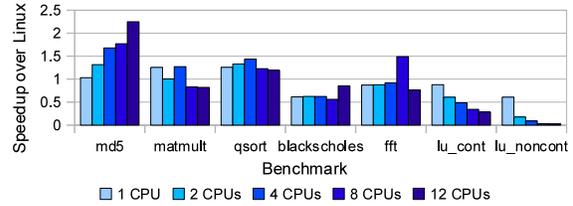}
\caption{Determinator performance relative to Linux
	on various parallel benchmarks.}
\label{fig-perf-bench}
\end{figure}

Figure~\ref{fig-perf-bench} shows the performance
of several shared-memory parallel benchmarks we ported,
relative to the same benchmarks running on
the 32-bit version of Ubuntu Linux 9.10.
The {\em md5} benchmark searches for an ASCII string
yielding a particular MD5 hash,
as in a brute-force password cracker;
{\em matmult} multiplies two $1024\times1024$ integer matrices;
{\em qsort} performs a recursive parallel quicksort on an integer array;
{\em blackscholes} is a financial benchmark from the PARSEC suite~\cite{
	bienia08characterization};
and {\em fft}, {\em lu\_cont}, and {\em lu\_noncont}
are Fast Fourier Transform and LU-decomposition benchmarks
from SPLASH-2~\cite{woo95splash2}.
We tested all benchmarks on a 2 socket $\times$ 6 core, 2.2GHz AMD Opteron PC.

Coarse-grained benchmarks
like {\em md5}, {\em matmult}, {\em qsort}, {\em blackscholes},
and {\em fft}
show performance comparable
with that of nondeterministic multithreaded execution under Linux.
The {\em md5} benchmark shows better scaling on Determinator than on Linux,
achieving a $2.25\times$ speedup over Linux on 12 cores.
We have not identified the precise cause of this speedup over Linux
but suspect scaling bottlenecks in Linux's thread system~\cite{behren03}.

Porting the {\em blackscholes} benchmark to Determinator
required no changes
as it uses deterministically scheduled pthreads
(Section~\ref{sec-dsched}).
The deterministic scheduler's quantization, however,
incurs a fixed performance cost of about 35\%
for the chosen quantum of 10 million instructions.
We could reduce this overhead by increasing the quantum,
or eliminate it by porting the benchmark
to Determinator's ``native'' parallel API.

The fine-grained {\em lu} benchmarks
show a higher performance cost,
indicating that Determinator's virtual memory-based approach
to enforcing determinism is not well-suited
to fine-grained parallel applications.
Future hardware enhancements might make determinism practical
for fine-grained parallel applications, however~\cite{devietti09dmp}.

\begin{figure}[t]
\centering
\includegraphics[width=0.48\textwidth]{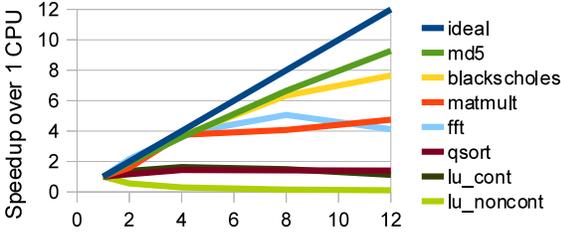}
\caption{Determinator parallel speedup over single-CPU performance
	on various benchmarks.}
\label{fig-perf-speedup}
\end{figure}

Figure~\ref{fig-perf-speedup} shows each benchmark's speedup
relative to single-threaded execution on Determinator.
The ``embarrassingly parallel'' {\em md5} and {\em blackscholes}
scale well,
{\em matmult} and {\em fft} level off after four processors
(but still perform comparably to Linux
as Figure~\ref{fig-perf-bench} shows),
and the remaining benchmarks scale poorly.

\begin{figure}[t]
\centering
\includegraphics[width=0.48\textwidth]{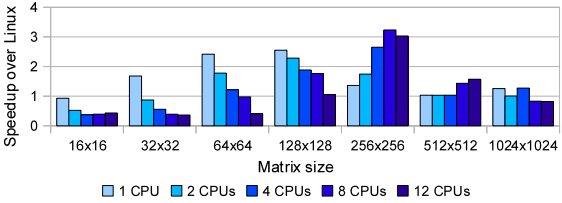}
\caption{Matrix multiply with varying matrix size.}
\label{fig-perf-matmult}
\end{figure}

\begin{figure}[t]
\centering
\includegraphics[width=0.48\textwidth]{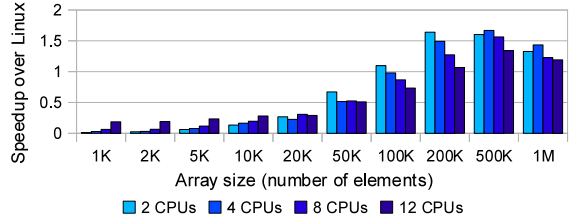}
\caption{Parallel quicksort with varying array size.}
\label{fig-perf-pqsort}
\end{figure}

To quantify further the effect of parallel interaction granularity
on deterministic execution performance,
Figures~\ref{fig-perf-matmult} and~\ref{fig-perf-pqsort}
show Linux-relative performance of {\em matmult} and {\em qsort},
respectively, for varying problem sizes.
With both benchmarks,
deterministic execution incurs a high performance cost
on small problem sizes requiring frequent interaction,
but on large problems Determinator is competitive with
and sometimes faster than Linux.

\subsection{Distributed Computing Performance}

While Determinator's rudimentary space migration (Section~\ref{sec-dist})
is far from providing a full cluster computing architecture,
we would like to test whether such a mechanism
can extend a deterministic computing model across nodes
with usable performance at least for some applications.
We therefore changed the {\em md5} and {\em matmult} benchmarks
to distribute workloads across a cluster
of up to 32 uniprocessor nodes via space migration.
Both benchmarks still run in a (logical) shared memory model via Snap/Merge.
\com{
We ran these benchmarks on a version of Determinator/PIOS
that a student in Yale's OS course implemented
and refined for better cross-node operation as his final class project.
}
Since we did not have a cluster on which we could run Determinator natively,
we ran it under QEMU~\cite{bellard05qemu},
on a cluster of
2 socket $\times$ 2 core, 2.4GHz Intel Xeon machines
running SuSE Linux 11.1.

\begin{figure}[t]
\centering
\includegraphics[width=0.40\textwidth]{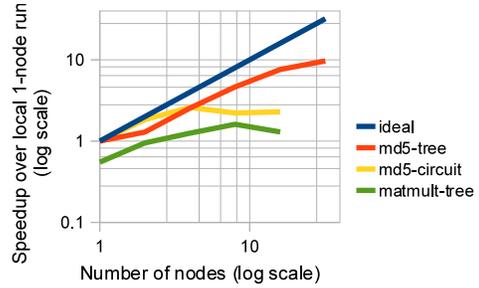}
\caption{MD5 benchmark on varying-size clusters.}
\label{fig-perf-mm-speedup}
\end{figure}

Figure~\ref{fig-perf-mm-speedup} shows parallel speedup under Determinator
relative to local single-node execution in the same environment,
on a log-log scale.
In {\em md5-circuit},
the master space acts like a traveling salesman,
migrating serially to each ``worker'' node to fork child processes,
then retracing the same circuit to collect their results.
The {\em md5-tree} variation forks workers recursively
in a binary tree:
the master space forks children on two nodes,
those children each fork two children on two nodes, etc.
The {\em matmult-tree} benchmark implements matrix multiply
with recursive work distribution as in {\em md5-tree}.

The ``embarrassingly parallel'' {\em md5-tree}
performs and scales well, but only with recursive work distribution.
Matrix multiply levels off at two nodes,
due to the amount of matrix data the kernel
transfers across nodes via its simplistic page copying protocol,
which currently performs no data streaming, prefetching, or delta compression.
The slowdown for 1-node distributed execution in {\em matmult-tree}
reflects the cost of transferring the matrix
to a (single) remote machine for processing.

\begin{figure}[t]
\centering
\includegraphics[width=0.40\textwidth]{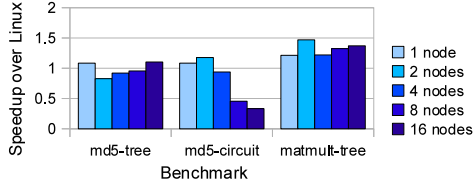}
\caption{Deterministic, shared-memory MD5 benchmark compared with a
	nondeterministic, distributed-memory Linux implementation.}
\label{fig-perf-mm-linux}
\end{figure}

Figure~\ref{fig-perf-mm-linux} shows that
the shared memory {\em md5-tree} and {\em matmult-tree} benchmarks,
running on Determinator,
perform comparably to nondeterministic, distributed-memory equivalents
running on Puppy Linux 4.3.1,
in the same QEMU environment.
The Determinator version of {\em md5}
is 63\% the size of the Linux version
(62 lines containing semicolons versus 99),
which uses remote shells to coordinate workers.
The Determinator version of {\em matmult}
is 34\% the size of its Linux equivalent (90 lines versus 263),
which passes data via TCP.

\com{
we compared Determinator against Puppy Linux
we compared against Linux under qemu. Tests were run on a group of 2 socket $\times$
2 core, 2.4GHz Intel Xeon machines, running SuSE Linux 11.1, kernel version 2.6.27.45.
PuppyLinux was used to run inside qemu because of the lightweight and the ease
of deployment. Version 4.3.1 of it comes with kernel version 2.6.30.5.

XXX block pwcrack overhead dominates for node $\geq$ 4 but linux scaled linearly.
Tree-like parallel structure still speeds up on 32 machines.

XXX matrix multiplication need data at least square root to the amount of
computation, for 1024$\times$1024 case, doesn't scale well over 8 machines.
(what about single machine?) (workload division method matters)

\begin{figure}[t]
\centering
\includegraphics[width=0.48\textwidth]{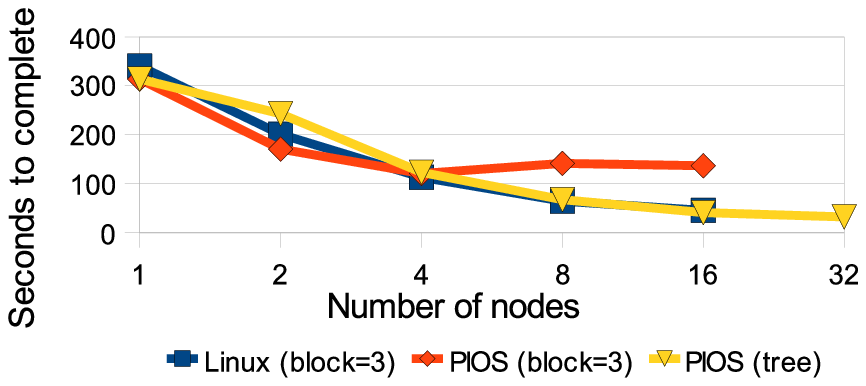}
\caption{Parallel MD5 cracking on varying number of nodes.}
\label{fig-perf-mm-pwcrack}
\end{figure}

\begin{figure}[t]
\centering
\includegraphics[width=0.48\textwidth]{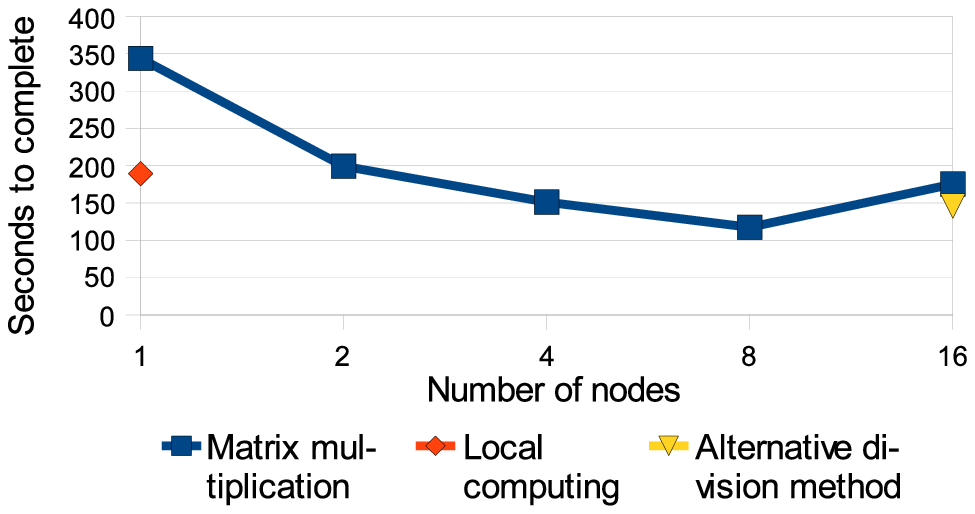}
\caption{1024$\times$1024 matrix multiply on varying number of nodes.}
\label{fig-perf-mm-matmult}
\end{figure}
}

\subsection{Implementation Complexity}

\begin{table}[t]
\centering
\begin{small}
\begin{tabular}{l|r|r}
			& Determinator	& PIOS		\\
Component		& Semicolons	& Semicolons	\\
\hline                                                  
Kernel core		& 2044		& 1847		\\
Hardware/device drivers	& 751		& 647		\\
User-level runtime	& 2952		& 1079		\\
Generic C library code	& 6948		& 394		\\
User-level programs	& 1797		& 1418		\\
\hline                                                  
Total			& 14,492	& 5385		\\
\end{tabular}
\end{small}
\caption{Implementation code size of the Determinator OS
	and of PIOS, its instructional subset.}
\label{tab-codesize}
\end{table}

To provide a feel for implementation complexity,
Table~\ref{tab-codesize} shows source code line counts for Determinator,
as well as its PIOS instructional subset,
counting only lines containing semicolons.
The entire system is less than 15,000 lines,
about half of which is generic C and math library code
needed mainly for porting Unix applications easily.

%% file: related.tex
\section{Related Work}
\label{sec-related}

The benefits of deterministic programming models
are well-known~\cite{lee06problem,bocchino09parallel}.
Recognizing these benefits,
parallel languages such as SHIM~\cite{
	edwards06shim,tardieu06scheduling,edwards08programming}
and DPJ~\cite{bocchino09parallel,bocchino09dpj}
enforce determinism at language level,
but cannot run legacy or multi-process parallel code.
Race detectors~\cite{engler03racerx,musuvathi08heisenbugs}
can detect heisenbugs in nondeterministic parallel programs,
but may miss heisenbugs resulting
from higher-level order dependencies~\cite{artho03high}.
Language extensions
can dynamically check determinism assertions in parallel code~\cite{
	sadowski09singletrack,burnim09asserting},
but heisenbugs may persist if the programmer omits an assertion.
Only a deterministic environment
prevents heisenbugs in the first place.

Application-level deterministic schedulers
such as DMP~\cite{devietti09dmp}, Grace~\cite{berger09grace},
and CoreDet~\cite{bergan10coredet}
instrument an application process
to isolate threads' memory accesses,
and run the threads on an artificial, deterministic execution schedule.
DMP and CoreDet isolate threads via code rewriting,
while Grace uses virtual memory techniques as in Determinator.
Since these schedulers run
in the same process as the application itself,
bugs or malicious code can violate determinism
by corrupting the scheduler,
as the authors acknowledge.
Determinator's kernel-enforced model
ensures repeatability of arbitrary code
in both multithreaded and multi-process computations.
Determinator's user-level runtime also develops deterministic versions
of OS abstractions such as shared file systems,
which lie outside the domain
of application-level deterministic schedulers.

DMP and Grace emulate sequential consistency~\cite{
	lamport79multi}
by running parallel tasks speculatively,
detecting read/write dependencies between tasks,
and re-executing tasks serially on detecting a dependency.
DMP-B~\cite{bergan10coredet} relaxes memory consistency
to optimize parallel execution,
but still emulates a nondeterministic programming model
where writes propagate between threads
at arbitrary points unpredictable to the developer.
Determinator combines ideas from early parallel Fortran systems~\cite{
	schwartz80burroughs,beltrametti88control}
with release consistency~\cite{
	gharachorloo90memory,carter91implementation,
	keleher92lazy,amza96treadmarks}
to develop a ``naturally deterministic'' programming model~\cite{
	ford10deterministic}.
In this model,
read/write conflicts do not exist (only write/write conflicts),
and shared memory or file changes
propagate among concurrent threads or processes
{\em only} at explicit synchronization points.
While focusing on this deterministic programming model,
Determinator's runtime can emulate nondeterministic models
via deterministic scheduling to run legacy parallel code.

\com{ XXX programming model precedent:
	Burroughs FMP (schwartz80burroughs)
	possibly definitive reference, but can't find:
		Lundstrom, A controllable MIMD architecture
	related, but not really suitable:
		Lundstrom, A Decentralized, Highly Concurrent Multiprocessor
		Schwartz, Ultracomputers

	spot-on description, but later:
	Myrias parallel computer (beltrametti88control)
}

Many techniques are available
for logging and replaying nondeterministic events
in parallel applications~\cite{
	curtis82bugnet,leblanc87debugging,feldman88igor,pan88supporting}.
SMP-ReVirt can log and replay a multiprocessor virtual machine~\cite{
	dunlap08execution},
supporting uses such as system-wide
intrusion analysis~\cite{dunlap02revirt,joshi05detecting}
and replay debugging~\cite{king05debugging}.
Logging a parallel system's nondeterministic events
is costly in performance and storage space, however,
and usually infeasible for ``normal-case'' execution.
Determinator demonstrates the feasibility
of providing system-enforced determinism
for normal-case execution,
without internal event logging,
while maintaining performance comparable with current systems
at least for coarse-grained parallel applications.
\com{
The performance and space costs of logging nondeterministic events
usually make replay usable only ``in the lab,'' however:
if a bug or intrusion manifests under deployment with logging disabled,
the event may not be subsequently reproducible.
In a deterministic environment, any event is reproducible
provided only that the original external inputs to the computation are logged.
}

Transactional memory (TM)~\cite
	{herlihy93transactional,shavit97software}
isolate threads' writes from each other
between transaction start and commit/abort.
TM offers no deterministic ordering between transactions, however:
like mutex locks,
transactions guarantee only atomicity, not determinism.


%% file: concl.tex
\section{Conclusion}
\label{sec-concl}

Determinator is only a first step towards
making deterministic execution readily available
and broadly usable for normal-case execution of parallel applications.
Nevertheless, our experiments suggest that,
with appropriate kernel and user-level runtime designs,
it is possible to provide system-enforced deterministic execution
efficiently at least for coarse-grained parallel applications,
both on a single multicore machine and across a cluster.

\com{
We have introduced a general approach
to combating timing channels in clouds
via provider-enforced deterministic execution,
without undermining the cloud's elasticity
through resource partitioning.
Preliminary results from our determinism-enforcing hypervisor
suggest that such a timing-hardened architecture
may be feasible and efficient at least for some applications,
but many questions remain.
Can such an architecture support
fine-grained parallel applications,
interactive Web applications,
transactional storage- or communication-intensive applications?
Can it offer cloud customers
a rich and convenient, yet efficient,
programming model in which to express such applications deterministically?
Can deterministic clouds reuse legacy software and operating systems?
Only further exploration will tell.
}

\com{
\subsection*{Acknowledgments}

Zhong
Ramki

funding?
}

%% file: main.bbl
\begin{thebibliography}{10}

\bibitem{hp94parisc}
{\em PA-RISC 1.1 Architecture and Instruction Set Reference Manual}.
\newblock Hewlett-Packard, third edition, Feb. 1994.

\bibitem{amza96treadmarks}
C.~Amza et~al.
\newblock {TreadMarks}: Shared memory computing on networks of workstations.
\newblock {\em IEEE Computer}, 29(2):18--28, Feb. 1996.

\bibitem{artho03high}
C.~Artho, K.~Havelund, and A.~Biere.
\newblock High-level data races.
\newblock In {\em \bibbrev{VVEIS}{Workshop on Verification and Validation of
  Enterprise Information Systems (VVEIS)}}, pages 82--93, Apr. 2003.

\bibitem{ford10determinating}
A.~Aviram and B.~Ford.
\newblock Determinating timing channels in statistically multiplexed clouds,
  Mar. 2010.
\newblock \url{http://arxiv.org/abs/1003.5303}.

\bibitem{ford10deterministic}
A.~Aviram and B.~Ford.
\newblock Deterministic consistency: A programming model for shared memory
  parallelism, Feb. 2010.
\newblock \url{http://arxiv.org/abs/0912.0926}.

\bibitem{bellard05qemu}
F.~Bellard.
\newblock {QEMU}, a fast and portable dynamic translator, Apr. 2005.

\bibitem{beltrametti88control}
M.~Beltrametti, K.~Bobey, and J.~R. Zorbas.
\newblock The control mechanism for the {Myrias} parallel computer system.
\newblock {\em Computer Architecture News}, 16(4):21--30, Sept. 1988.

\bibitem{bergan10coredet}
T.~Bergan, O.~Anderson, J.~Devietti, L.~Ceze, and D.~Grossman.
\newblock {CoreDet}: A compiler and runtime system for deterministic
  multithreaded execution.
\newblock In {\em 15th \bibbrev{ASPLOS}{international Conference on
  Architectural Support for Programming Languages and Operating Systems
  (ASPLOS)}}, Mar. 2010.

\bibitem{berger09grace}
E.~D. Berger, T.~Yang, T.~Liu, and G.~Novark.
\newblock {Grace}: Safe multithreaded programming for {C/C++}.
\newblock In {\em OOPSLA}, Oct. 2009.

\bibitem{bershad95extensibility}
B.~N. Bershad et~al.
\newblock Extensibility, safety and performance in the {SPIN} operating system.
\newblock In {\em 15th \bibbrev{SOSP}{ACM Symposium on Operating System
  Principles}}, 1995.

\bibitem{bienia08characterization}
C.~Bienia, S.~Kumar, J.~P. Singh, and K.~Li.
\newblock The {PARSEC} benchmark suite: Characterization and architectural
  implications.
\newblock In {\em 17th International Conference on Parallel Architectures and
  Compilation Techniques}, October 2008.

\bibitem{openmp08}
O.~A.~R. Board.
\newblock {OpenMP} application program interface version 3.0, May 2008.
\newblock \url{http://www.openmp.org/mp-documents/spec30.pdf}.

\bibitem{bocchino09parallel}
R.~L. Bocchino~Jr., V.~S. Adve, S.~V. Adve, and M.~Snir.
\newblock Parallel programming must be deterministic by default.
\newblock In {\em \bibconf[1st]{HotPar}{Workshop on Hot Topics in Parallelism
  (HotPar '09)}}. Mar. 2009.

\bibitem{bocchino09dpj}
R.~L. {Bocchino Jr.}, V.~S. Adve, D.~Dig, S.~V. Adve, S.~Heumann,
  R.~Komuravelli, J.~Overbey, P.~Simmons, H.~Sung, and M.~Vakilian.
\newblock A type and effect system for {Deterministic Parallel Java}.
\newblock Oct. 2009.
\newblock \url{http://dpj.cs.uiuc.edu/DPJ/Publications_files/paper_1.pdf}.

\bibitem{bressoud96hypervisor}
T.~C. Bressoud and F.~B. Schneider.
\newblock Hypervisor-based fault-tolerance.
\newblock {\em \bibbrev{TOCS}{ACM Transactions on Computer Systems}},
  14(1):80--107, Feb. 1996.

\bibitem{burnim09asserting}
J.~Burnim and K.~Sen.
\newblock Asserting and checking determinism for multithreaded programs.
\newblock In {\em \bibbrev{FSE}{ACM SIGSOFT Symposium on the Foundations of
  Software Engineering}}, Aug. 2009.

\bibitem{carter91implementation}
J.~B. Carter, J.~K. Bennett, and W.~Zwaenepoel.
\newblock Implementation and performance of munin.
\newblock In {\em 13th \bibbrev{SOSP}{{ACM} Symposium on Operating Systems
  Principles (SOSP)}}, Oct. 1991.

\bibitem{castro99practical}
M.~Castro and B.~Liskov.
\newblock Practical byzantine fault tolerance.
\newblock In {\em 3rd \bibbrev{OSDI}{USENIX Symposium on Operating Systems
  Design and Implementation (OSDI)}}, pages 173--186, Feb. 1999.

\bibitem{tzicker99integrating}
T.~Chiueh, G.~Venkitachalam, and P.~Pradhan.
\newblock Integrating segmentation and paging protection for safe, efficient
  and transparent software extensions.
\newblock In {\em 17th \bibbrev{SOSP}{ACM Symposium on Operating System
  Principles}}, pages 140--153, Dec. 1999.

\bibitem{choi98deterministic}
J.-D. Choi and H.~Srinivasan.
\newblock Deterministic replay of {Java} multithreaded applications.
\newblock In {\em SPDT '98: Proceedings of the SIGMETRICS symposium on Parallel
  and distributed tools}, pages 48--59. 1998.

\bibitem{curtis82bugnet}
R.~S. Curtis and L.~D. Wittie.
\newblock {BugNet}: A debugging system for parallel programming environments.
\newblock In {\em 3rd \bibbrev{ICDCS}{International Conference on Distributed
  Computing Systems}}, pages 394--400, Oct. 1982.

\bibitem{devietti09dmp}
J.~Devietti, B.~Lucia, L.~Ceze, and M.~Oskin.
\newblock {DMP}: Deterministic shared memory multiprocessing.
\newblock In {\em 14th \bibbrev{ASPLOS}{international Conference on
  Architectural Support for Programming Languages and Operating Systems
  (ASPLOS)}}, Mar. 2009.

\bibitem{dunlap02revirt}
G.~W. Dunlap, S.~T. King, S.~Cinar, M.~A. Basrai, and P.~M. Chen.
\newblock {ReVirt}: Enabling intrusion analysis through virtual-machine logging
  and replay.
\newblock In {\em 5th \bibbrev{OSDI}{USENIX Symposium on Operating Systems
  Design and Implementation}}, Dec. 2002.

\bibitem{dunlap08execution}
G.~W. Dunlap, D.~G. Lucchetti, M.~A. Fetterman, and P.~M. Chen.
\newblock Execution replay for multiprocessor virtual machines.
\newblock In {\em \bibbrev{VEE}{Virtual Execution Environments (VEE)}}, Mar.
  2008.

\bibitem{edwards06shim}
S.~A. Edwards and O.~Tardieu.
\newblock Shim: A deterministic model for heterogeneous embedded systems.
\newblock {\em IEEE Transactions on Very Large Scale Integration (VLSI)
  Systems}, 14(8):854--867, Aug. 2006.

\bibitem{edwards08programming}
S.~A. Edwards, N.~Vasudevan, and O.~Tardieu.
\newblock Programming shared memory multiprocessors with deterministic
  message-passing concurrency: Compiling {SHIM} to {Pthreads}.
\newblock In {\em \bibbrev{DATE}{Design, Automation, and Test in Europe}}, Mar.
  2008.

\bibitem{engler03racerx}
D.~Engler and K.~Ashcraft.
\newblock {RacerX}: effective, static detection of race conditions and
  deadlocks.
\newblock In {\em 19th \bibbrev{SOSP}{{ACM} Symposium on Operating Systems
  Principles (SOSP)}}, Oct. 2003.

\bibitem{feldman88igor}
S.~I. Feldman and C.~B. Brown.
\newblock {IGOR}: A system for program debugging via reversible execution.
\newblock In {\em Workshop on Parallel \& Distributed Debugging}, pages
  112--123, May 1988.

\bibitem{ford96microkernels}
B.~Ford, M.~Hibler, J.~Lepreau, P.~Tullmann, G.~Back, and S.~Clawson.
\newblock Microkernels meet recursive virtual machines.
\newblock In {\em 2nd \bibbrev{OSDI}{USENIX Symposium on Operating Systems
  Design and Implementation (OSDI)}}, pages 137--151, 1996.

\bibitem{garfinkel07compatibility}
T.~Garfinkel, K.~Adams, A.~Warfield, and J.~Franklin.
\newblock Compatibility is not transparency: {VMM} detection myths and
  realities.
\newblock In {\em \bibbrev{HotOS-XI}{11th Workshop on Hot Topics in Operating
  Systems}}, May 2007.

\bibitem{gharachorloo90memory}
K.~Gharachorloo, D.~Lenoski, J.~Laudon, P.~Gibbons, A.~Gupta, and J.~Hennessy.
\newblock Memory consistency and event ordering in scalable shared-memory
  multiprocessors.
\newblock In {\em 17th \bibbrev{ISCA}{International Symposium on Computer
  Architecture}}, pages 15--26, May 1990.

\bibitem{goldberg96secure}
I.~Goldberg, D.~Wagner, R.~Thomas, and E.~A. Brewer.
\newblock A secure environment for untrusted helper applications.
\newblock In {\em 6th USENIX Security Symposium}, 1996.

\bibitem{haeberlen07peerreview}
A.~Haeberlen, P.~Kouznetsov, and P.~Druschel.
\newblock {PeerReview}: Practical accountability for distributed systems.
\newblock In {\em 21st \bibbrev{SOSP}{{ACM} Symposium on Operating Systems
  Principles (SOSP)}}, Oct. 2007.

\bibitem{halstead85multilisp}
R.~H. {Halstead, Jr.}
\newblock {Multilisp}: A language for concurrent symbolic computation.
\newblock {\em \bibbrev{TOPLAS}{ACM Transactions on Programming Languages and
  Systems}}, 7(4):501--538, Oct. 1985.

\bibitem{herlihy93transactional}
M.~Herlihy and J.~E.~B. Moss.
\newblock Transactional memory: Architectural support for lock-free data
  structures.
\newblock In {\em 20th \bibbrev{ISCA}{International Symposium on Computer
  Architecture}}, pages 289--300, May 1993.

\bibitem{joshi05detecting}
A.~Joshi, S.~T. King, G.~W. Dunlap, and P.~M. Chen.
\newblock Detecting past and present intrusions through vulnerability-specific
  predicates.
\newblock In {\em SOSP '05: Proceedings of the twentieth ACM symposium on
  Operating systems principles}, pages 91--104. 2005.

\bibitem{mit-jos}
F.~Kaashoek et~al.
\newblock 6.828: Operating system engineering.
\newblock \url{http://pdos.csail.mit.edu/6.828/}.

\bibitem{kahn74semantics}
G.~Kahn.
\newblock The semantics of a simple language for parallel programming.
\newblock In {\em Information Processing}, pages 471--475. 1974.

\bibitem{keleher92lazy}
P.~Keleher, A.~L. Cox, and W.~Zwaenepoel.
\newblock Lazy release consistency for software distributed shared memory.
\newblock In {\em \bibbrev{ISCA}{13th International Symposium on Computer
  Architecture}}, pages 13--21, May 1992.

\bibitem{king05debugging}
S.~T. King, G.~W. Dunlap, and P.~M. Chen.
\newblock Debugging operating systems with time-traveling virtual machines.
\newblock In {\em \bibbrev{USENIX}{USENIX Annual Technical Conference}}, pages
  1--15, Apr. 2005.

\bibitem{lamport79multi}
L.~Lamport.
\newblock How to make a multiprocessor computer that correctly executes
  multiprocess programs.
\newblock {\em {IEEE} Transactions on Computers}, 28(9):690--691, Sept. 1979.

\bibitem{leblanc87debugging}
T.~J. Leblanc and J.~M. Mellor-Crummey.
\newblock Debugging parallel programs with instant replay.
\newblock {\em IEEE Transactions on Computers}, C-36(4):471--482, Apr. 1987.

\bibitem{lee06problem}
E.~Lee.
\newblock The problem with threads.
\newblock {\em Computer}, 39(5):33--42, May 2006.

\bibitem{lu08learning}
S.~Lu, S.~Park, E.~Seo, and Y.~Zhou.
\newblock Learning from mistakes --- a comprehensive study on real world
  concurrency bug characteristics.
\newblock In {\em 13th \bibbrev{ASPLOS}{international Conference on
  Architectural Support for Programming Languages and Operating Systems
  (ASPLOS)}}, pages 329--339, Mar. 2008.

\bibitem{musuvathi08heisenbugs}
M.~Musuvathi, S.~Qadeer, T.~Ball, and G.~Basler.
\newblock Finding and reproducing heisenbugs in concurrent programs.
\newblock In {\em Proceedings of the 8th USENIX Symposium on Operating System
  Design and Implementation (OSDI '08)}, pages 267--280. 2008.

\bibitem{pan88supporting}
D.~Z. Pan and M.~A. Linton.
\newblock Supporting reverse execution of parallel programs.
\newblock In {\em \bibbrev{PADD '88}{Workshop on Parallel and Distributed
  Debugging (PADD)}}, pages 124--129. 1988.

\bibitem{parker83detection}
D.~S. {Parker, Jr.} et~al.
\newblock Detection of mutual inconsistency in distributed systems.
\newblock {\em IEEE Transactions on Software Engineering}, SE-9(3), May 1983.

\bibitem{sadowski09singletrack}
C.~Sadowski, S.~N. Freund, and C.~Flanagan.
\newblock {SingleTrack}: A dynamic determinism checker for multithreaded
  programs.
\newblock In {\em 18th \bibbrev{ESOP}{European Symposium on Programming}}, Mar.
  2009.

\bibitem{schneider90implementing}
F.~B. Schneider.
\newblock Implementing fault-tolerant services using the state machine
  approach: A tutorial.
\newblock Technical Report 86-800, Cornell University, Jan. 1990.

\bibitem{schwartz80burroughs}
J.~T. Schwartz.
\newblock The burroughs {FMP} machine, Jan. 1980.
\newblock Ultracomputer Note \#5.

\bibitem{shavit97software}
N.~Shavit and D.~Touitou.
\newblock Software transactional memory.
\newblock {\em Distributed Computing}, 10(2):99--116, Feb. 1997.

\bibitem{tardieu06scheduling}
O.~Tardieu and S.~A. Edwards.
\newblock Scheduling-independent threads and exceptions in {SHIM}.
\newblock In {\em \bibbrev{EMSOFT}{6th Conference on Embedded Software}}, pages
  142--151, Oct. 2006.

\bibitem{terry95managing}
D.~B. Terry et~al.
\newblock Managing update conflicts in {Bayou}, a weakly connected replicated
  storage system.
\newblock In {\em 15th \bibbrev{SOSP}{ACM Symposium on Operating System
  Principles}}, 1995.

\bibitem{behren03}
R.~von Behren, J.~Condit, F.~Zhou, G.~C. Necula, and E.~Brewer.
\newblock Capriccio: {S}calable threads for internet services.
\newblock In {\em {SOSP}'03}.

\bibitem{walker83locus}
B.~Walker, G.~Popek, R.~English, C.~Kline, and G.~Thiel.
\newblock The {LOCUS} distributed operating system.
\newblock {\em SIGOPS Operating Systems Review}, 17(5), Oct. 1983.

\bibitem{woo95splash2}
S.~C. Woo, M.~Ohara, E.~Torrie, J.~P. Singh, and A.~Gupta.
\newblock The {SPLASH-2} programs: Characterization and methodological
  considerations.
\newblock In {\em 22nd \bibbrev{ISCA}{International Symposium on Computer
  Architecture}}, pages 24--36, June 1995.

\end{thebibliography}
